\documentstyle[12pt,twoside,fleqn,espcrc1]{article}

\newcommand{\AmS}{{\protect\the\textfont2
  A\kern-.1667em\lower.5ex\hbox{M}\kern-.125emS}}
\input{epsf}

\begin{document}

\noindent
\hspace{8.0cm} University of Bern preprint { BUHE-2000-1}
\\

\hspace{12.0cm} 7 April 2000\\

\vspace{0.5cm}

 \begin{center}
{ \large Charm in nuclear reactions at $\sqrt{s}$ = 17 and 19 GeV}\footnote{Work
supported by the Swiss National Science Foundation.}
\vspace{0.1cm}

Sonja Kabana
\vspace{0.1cm}

 Laboratory for High Energy Physics, University of Bern,
\\
 Sidlerstrasse 5, 3012 Bern, Switzerland
\\
E-mail: sonja.kabana@cern.ch
\\
 \end{center}

\vspace{1.cm}

\noindent
\begin{center}
{\bf Abstract}
\end{center}

\noindent
Consequences resulting from the $D \overline{D}$ excess derived
indirectly by the NA50 experiment in S+U and Pb+Pb collisions at 
$\sqrt{s}$=19, 17 GeV, relevant for the identification of the
QCD phase transition in these collisions, are  discussed. 
The dependence of open and closed charm yields in
Pb+Pb collisions  on the number of participating
nucleons (N) indicates non thermal charm production and
$J/\Psi$ dissociation, stronger than the absorption seen in any other elementary hadron.
The $J/\Psi$ in central Pb+Pb collisions
 could originate dominantly from $c \overline{c}$ 
 pair coalescence out of a hadronizing quark and gluon environment.
Furthermore, the $J/\Psi$ appears to be suppressed  in S+U collisions
at $\sqrt{s}$=19 GeV, as opposed to current interpretations.
A significant change in the $(J/\Psi)/ D \overline{D}$ ratio 
as well as in the number density of kaons is
 observed above energy density $\epsilon$ $\sim$ 1 GeV/fm$^3$,
 suggesting  a change of phase  at this energy density,
and underlining the importance of direct open charm measurements.

\section{Introduction}

\noindent
Quantum chromodynamics (QCD) on 
 the lattice predicts a phase transition of
confined hadronic matter into deconfined quark and gluon matter 
(called the Quark Gluon Plasma state -- QGP)
at a critical temperature $T_c$ $\sim$ 150 MeV, respectively
at energy density $\epsilon_c$ $\sim$ 1 GeV/fm$^3$ \cite{lattice}.
The order of the transition is parameter dependent \cite{order}.
An investigation of relevant observables in heavy ion collisions
 \cite{horst}
 as a function of energy and/or the impact parameter\footnote{Provided 
that there is a unique assignement between impact parameter and QGP 
phase transition.}
of the collision
could reveal this transition, through a discontinuous behaviour
 of many QGP signatures at the transition point.
\\

\noindent
An anomalous suppression of the $J/\Psi$ meson predicted to be a 
signature of Quark Gluon Plasma formation \cite{jpsiprediction} has been
measured  to occur in the ratio of the $J/\Psi$ over the
 Drell Yan (DY) process 
in Pb+Pb collisions at $\sqrt{s}$=17 GeV investigated  as a function
of transverse energy ($E_T$) \cite{cernep_2000_13}.
In S+U collisions at $\sqrt{s}$=19 GeV and in the most peripheral Pb+Pb
collisions, the $(J/\Psi)/DY$ ratio agrees
with expectations  \cite{cernep_2000_13}.
The ratio $(J/\Psi)/DY$ is relevant for the investigation of the
$J/\Psi$ suppression, under the assumption that
$J/\Psi$  production in these collisions is a hard process.
\\

\noindent
Recent 
measurements of the dimuon invariant mass m($\mu^+ \mu^-$) spectrum
between the $\phi$ and the $J/\Psi$ mass (Intermediate Mass Region=IMR)
 revealed a dimuon enhancement 
above expectation, 
which is increasing with the number of nucleons participating
in the collision (N) \cite{cernep_2000_12}.
This enhancement can be understood as due to an excess of
$D \overline{D}$ production\footnote{With '$D \overline{D}$' we denote 
the number of $D$ and $\overline{D}$ hadrons which were simultaneously 
 found within the acceptance of the NA50 experiment
( $D \overline{D} \ = \ (D + \overline{D})_{acc}/2$ ).} 
as suggested by several features of the data, e.g. the
 shape of the mass, rapidity, angular and transverse
momentum distributions of the dimuons \cite{cernep_2000_12}.
The interpretation of the IMR enhancement as due to open charm is not 
unique though, because the open charm was extracted through
a fit to the dimuon continuum and the $D\overline{D}$ meson signal was
not directly identified.
Recent work suggesting that the seen enhancement could be due
to rescattering of D mesons in nuclear matter \cite{wang}
is not supported by the data \cite{soave_phd}.
An other possible interpretation is that the IMR excess
 could be due  to thermal dimuons \cite{thermal_dimuons}.
\noindent
Non-perturbative effects are known to play a role
in heavy flavour hadron production in elementary reactions
showing up in deviations of data from
perturbative QCD calculations \cite{mink_private,MINK}.
Based on this fact, one could expect that theoretical investigation
of non-perturbative effects and different reaction
dynamics as in the plasma phase may result 
in  an enhancement of open charm production
in nuclear reactions over perturbative QCD expectations.
\\

\noindent
If  the total charm produced in heavy ion collisions
 indeed deviates from the perturbative QCD expectations for a hard process as
suggested by the NA50 data,
it  follows that the $(J/\Psi)/DY$ ratio
is not the  proper 
 quantity for the search for
the $J/\Psi$ suppression as signature of  
Quark Gluon Plasma formation in nuclear reactions.
It is only the ratio $(J/\Psi)$/(total $c \overline{c}$) that
matters.
We therefore investigate here first the dependence 
of the $J/\Psi$ and the $D\overline{D}$ yields per collision
on N.
We further investigate the dependence of the
 $(J/\Psi)/D\overline{D}$ ratio on N and on the length of the 
nuclear matter traversed by the $J/\Psi$,
as well as the dependence of both charm and strangeness production on
the initial energy density reached in the collision.

\section{N dependence of open and closed charm yields in Pb+Pb collisions
at $\sqrt{s}$=17 GeV}

\subsection{N dependence of the Drell Yan yield}

\vspace{0.4cm}
\noindent
{\bf Calculation details}
\vspace{0.2cm}

\noindent
The dependence of the Drell Yan (DY) yield per nucleus-nucleus
collision in arbitrary units
 produced in Pb+Pb collisions at $\sqrt{s}$=17 GeV per N+N collision,
on the transverse energy of the collision
 has been measured by the NA50
collaboration (figure 7 in \cite{cernep_99_13}).
In this figure it is shown that the theoretically expected
DY yield, assuming DY production is a hard process, 
does partly reproduce the measured one;
the deviations at low transverse energy
 are understood
to result from properties of the lead (Pb) nucleus, in particular from 
the different radii of the proton and neutron distributions 
in Pb \cite{gerschel_private}.
It is therefore justified to use the theoretically
calculated $E_T$ dependence of DY yield per collision, to estimate
the dependence of $J/\Psi$ and $D\overline{D}$ yields on the number of
participating nucleons in the collision.
\\

\noindent
However, since the deviations of the very low statistics measurement of
DY yield at high $E_T$, from the theoretical
estimated DY yield  seen in figure 7 of
\cite{cernep_99_13} cannot be understood,
it would be important to {\it measure} the
DY yield per collision with high statistics in the
high $E_T$ region.
In this way the last drop in the $(J/\Psi)/DY$ ratio 
above N $\sim$ 360 \cite{cernep_2000_13}
can be experimentally verified.
\\

\subsection{N dependence of the $D\overline{D}$ yield}

\vspace{0.5cm}

\vspace{0.4cm}
\noindent
{\bf Calculation details}
\vspace{0.2cm}

\noindent
The NA50 collaboration
 observed an excess (E)
 of the measured over the expected $D\overline{D}/DY$ ratio 
in S+U and Pb+Pb collisions at $\sqrt{s}$=19, 17 GeV, 
which increases with the number of participants N
(figure 12 and table 4 in \cite{cernep_2000_12}).
If we fit the S+U and Pb+Pb E points of the above figure to a function 
$f=c \cdot N^{\alpha}$,  we find that
the excess is increasing with N as
 $N^{(\alpha=0.45 \pm 0.11)}$ ($\chi^2$/Degrees Of Freedom=1.7, DOF=7).
The $N$ dependence of the 
excess E of the $D\overline{D}/DY$ production in S+U collisions at 
$\sqrt{s}$=19 GeV and Pb+Pb collisions at $\sqrt{s}$=17 GeV 
 over expectations,
reflects the $N$ dependence of the $D\overline{D}/DY$ ratio.
This results from the fact, that all other quantities involved in
the definition of E \cite{cernep_2000_12,soave_phd},
 do not depend on $N$.
Therefore the $N$ dependence of the $D\overline{D}$
production yield is given by the $N$ dependence
  of the quantity 

\begin{equation}
n_{D\overline{D}} = E * n_{DY} \sim (D\overline{D} /DY) * n_{DY} 
\end{equation}

\noindent
where $n_{D\overline{D}}$, $n_{DY}$ denote the yields of $D\overline{D}$ 
 and DY per collision in arbitrary units.
The arbitrary units are due to the fact that NA50 did not published absolute
 yields per collision of the $J/\Psi$, $DY$ and 
$D \overline{D}$ separately, corrected for losses due to
e.g. acceptance,  as a function of N, $E_T$.
We suggest that it would be important to do so.
 \\

\noindent
The DY yield used for the above calculation
 has been extracted from the theoretical curve shown in
 figure 7 in \cite{cernep_99_13},
at the transverse energy
 ($E_T$) points in which the $D \overline{D}$ excess factor E
 has been measured.
\noindent
 The $E_T$ points corresponding to the excess factor E
 were extracted, by interpolating between the $E_T$ values given in 
table 1 of \cite{cernep_99_13}
 as a function of the mean impact parameter,
at the values of mean impact parameter for which
the factor E has been measured (listed in table 1 of \cite{cernep_2000_12}).
For the most central and the most peripheral points, for which no
mean b are given in table 1 of \cite{cernep_2000_12}, we estimated
the values of b, from the values of $N$ as a function of
  b for Pb+Pb collisions calculated by \cite{jyo_private,jyo_paper}.
These calculations \cite{jyo_private} agree 
with the values (N,b) estimated by NA50, when compared
in their common range. \\

\noindent
Though 
the $D \overline{D}$ measured by NA50 represents the joint
probability  that a $D $ and a $\overline{D}$ are both found 
in the experimental acceptance, the N dependence of it,
is expected to be the same or similar to the N dependence of the sum 
$D+ \overline{D}$, respectively of the $D$, and of the 
$\overline{D}$ hadrons.
It is however possible that the above N-dependences  deviate
from eachother, in case of a higher than 1
charm pair multiplicity per event with charm
(see \cite{marek_mult} for a numerical
 estimation of the charm multiplicity per event in central
Pb+Pb collisions at 158 A GeV).
Therefore, for clarity, the N dependence of the total extrapolated
$D+\overline{D}$ yield in nuclear collisions
 should be estimated including acceptance corrections by NA50.

\vspace{0.4cm}
\noindent
{\bf Results and discussion}
\vspace{0.2cm}

\noindent
The resulting $D\overline{D}$ yield in arbitrary units
 (figure \ref{ddbaryield}) 
increases as $N^{ (\alpha=1.70 \pm 0.12) }$
($\chi^2/DOF$ = 2.5, DOF=7).
This N behaviour indicates that $D\overline{D}$ production in Pb+Pb
collisions at $\sqrt{s}$=17 GeV,
 did not establish yet equilibrium,
in which case a proportionality with $N$ --assuming N to be 
proportional to the volume of the source
\footnote{
The assumption N $\sim$ V, is based on the observation that the freeze out 
volume V of the particle source is 
found to be proportional to N \cite{na52_centr}.}--
 is expected ($\alpha$= 1).
This appears justified because
the temperature in the collision zone --assuming local thermalization of
light particles--
drops with time and the mean temperature expected
 to be reached in these collisions of the order $\sim$ $10^2$ MeV,
 is much lower than the mass of charm quarks and/or charmed hadrons. 
\\

\subsection{N dependence of the $J/\Psi$ yield}
\vspace{0.5cm}

\vspace{0.4cm}
\noindent
{\bf Calculation details}
\vspace{0.2cm}

\noindent
In the following we estimate the $J/\Psi$ yield per 
collision as a function
of $N$, at the same $N$ values where the $D\overline{D}$
 was measured.
The $N$ dependence of the $J/\Psi$  yield per collision is 
given by the N dependence of the quantity :

\begin{equation}
n_{J/\Psi} = ( (J/\Psi)/ DY) * n_{DY} 
\end{equation}

\noindent
 where $n_{J/\Psi}$, $n_{DY}$ denote the yields of $J/\Psi$ and DY
in arbitrary common units.
The $(J/\Psi)/DY$ values have been extracted from  figure 4 of
\cite{cernep_2000_13} at the $E_T$ values where the E factor
has been measured, interpolating between the different points.
\\

\vspace{0.4cm}
\noindent
{\bf Results and discussion}
\vspace{0.2cm}

\noindent
The resulting $J/\Psi$ yield per collision produced in Pb+Pb collisions 
in arbitrary units (figure \ref{jpsiyield})
increases like $N^{ (\alpha= 0.70 \pm 0.04) }$ ($\chi^2/DOF$=1.43, DOF=7).
This N dependence indicates an increasing $J/\Psi$ dissociation 
with higher centrality.
 The strength of this dissociation as measured by the $\alpha$
 parameter, is higher for the $J/\Psi$ as compared to any other 
hadron\footnote{Deuterons have an even smaller 
$\alpha$ parameter, but they are not elementary hadrons and
 are weekly bound (see discussion in
\cite{na52_centr}).}  
 produced in these collisions for example as compared to
antiprotons.
For  the latter, a large annihilation cross section
with baryons is expected 
and there is indeed experimental evidence that they are 
absorbed with increasing centrality in Pb+Pb collisions
   ($\alpha$($\overline{p}$)=0.80 $\pm$ 0.04,
 ($\chi^2/DOF$=1.0, $DOF$=3)
   at y=3.7, $p_T$=0 \cite{na52_centr}\footnote{We extracted here the 
 $\alpha$ parameter for  $\overline{p}$,
 after quadratically adding the statistical and
a 5\% systematic error.}).
\\
\noindent
The $J/\Psi$ multiplicity as a function of N
extracted with an other method \cite{gerschel_paper},
 agrees  with the here presented results within the errors.
\\

\section{The $(J/\Psi)/D\overline{D}$ ratio in nuclear collisions}

\vspace{0.5cm}

\subsection{The N dependence of the $(J/\Psi)/D\overline{D}$ ratio in nuclear collisions}
\vspace{0.5cm}

\vspace{0.4cm}
\noindent
{\bf Calculation details}
\vspace{0.2cm}

\noindent
Assuming that the IMR excess is  due partly or solely to open charm,
 allows us to search for an anomalous suppression of
$J/\Psi$ as compared to the  open charm production.
The $N$ dependence of the $(J/\Psi)/D\overline{D}$ ratio in
Pb+Pb and S+U collisions at $\sqrt{s}$ of 17 and 19 GeV, estimated as:
\\

\begin{equation}
(J/\Psi)/D\overline{D} \sim
( (J/\Psi)/DY)  / (D\overline{D}/DY)
\sim 
( (J/\Psi) /DY)  / E 
\end{equation}

\noindent
in arbitrary units due to the E factor in equation (3), is a decreasing function of $N$ 
(figure \ref{jpsitoddbar_vs_n_pbpb_su}).
The $(J/\Psi)/DY$ in S+U collisions was taken from 
\cite{fleuret,phys_lett_b410_97_337}.
Note that possible deviations of the DY yield from its theoretical calculation
(as seen in figure 7 in \cite{cernep_99_13}),
do not drop out in the 
$( (J/\Psi)/DY)/ (D \overline{D}/DY)$ ratio shown here, because
the $D \overline{D}/DY$ --unlike the $(J/\Psi)/DY$--
 was  calculated by NA50
not using the minimum bias theoretical DY yield values
but the measured ones.
\\

\noindent
In order to show the influence of the very last drop of $(J/\Psi)/DY$, on the 
$(J/\Psi)/D \overline{D}$ ratio, the $(J/\Psi)/DY$ ratio
divided by $N^{ 0.45 \pm 0.11}$ is plotted as a function of N
in figure \ref{jpsitoddbar_to_nto045}.
This quantity resembles the 
$(J/\Psi)/D \overline{D}$ ratio

\begin{equation}
(J/\Psi)/D \overline{D} 
\sim
((J/\Psi)/DY)/ N^{ 0.45 \pm 0.11}
\end{equation}

\noindent
in arbitrary units,
 because $N^{ 0.45 \pm 0.11}$ is the
found N dependence of the $D \overline{D}/DY$ ratio.
The open points of figure \ref{jpsitoddbar_to_nto045}
are extracted from the 'minimum bias analysis'
 results of figure 4 in \cite{cernep_2000_13}\footnote{'Minimum
 bias analysis' in NA50, means that
the DY for the $(J/\Psi)/DY$  ratio,
was determined using the theoretically estimated DY yield per collision
as a function of $E_T$
 and the measured $dN / dE_T$ vs $E_T$
 spectrum of minimum bias trigger events (see \cite{cernep_99_13}).}.
The closed points show the $((J/\Psi)/DY)/ N^{0.45} $ 
calculated here, at the N values
where the $D \overline{D}$ excess factor was measured.

\vspace{0.4cm}
\noindent
{\bf Results and discussion}
\vspace{0.2cm}

\noindent
The $(J/\Psi)/D\overline{D}$ ratio in Pb+Pb collisions
found as shown in equation (3),
decreases with $N$ as $N^{ (\alpha= -0.79 \pm 0.14) }$ ($\chi^2/DOF$ = 3.3,
$DOF$=7),
respectively  like $N^{ (\alpha= -1.17 \pm 0.14) }$ ($\chi^2/DOF$=1.54, DOF=6)
when the first point is not fitted\footnote{The first point
of the  $D\overline{D}/DY$ enhancement factor E
 lyes significantly above the $N^{\alpha}$ function
fit to the E distribution (figure 12 in \cite{cernep_2000_12}).}.
The $(J/\Psi)/D\overline{D}$ ratio in S+U collisions 
(figure \ref{jpsitoddbar_vs_n_pbpb_su}) 
decreases with $N$ as $N^{ (\alpha= -0.62 \pm 0.22) }$ ($\chi^2/DOF$=0.69,
$DOF$=3). 
The $(J/\Psi)/D \overline{D}$ found as shown in equation (4)
 when fitted to the function $N^{\alpha}$
 until N=380, gives
$N^{ (\alpha=-1.07 \pm 0.07) }$, ($\chi/DOF$=0.93, DOF=7). 
\\

\noindent
If the $J/\Psi$ is completely dissociated in a quark gluon plasma
and is  formed later mainly through $c$ and $\overline{c}$
 quark coalescence, we expect that the N dependence of the
ratio  $(J/\Psi)/D \overline{D}$ 
--rather than the $(J/\Psi)/(D \overline{D})^2$--
 reflects the N dependence of the volume of the charm environment 
  \cite{mink_private}.
This is due to the expectation that because of the very low
cross section of charm production at these energies,
 there is most often just one
$c \overline{c}$ pair per event containing charm,  whatever N.
Then
 the probability to form a $J/\Psi$ from coalescence is
proportional to 
$(J/\Psi)/D \overline{D}$ and
inversely proportional to the volume of the particle source --made up by
$u \overline{u} d \overline{d} s \overline{s}$ quarks and gluons-- within which
the $c$ and $\overline{c}$ quarks scatter. 
Assuming this volume is proportional to N (see footnote 3), one would expect
that $(J/\Psi)/D \overline{D}$ decreases as $N^{-1}$,
as actually observed.
\\

\noindent
In this case, one can use the $(J/\Psi)/D\overline{D}$
 ratio to extract the absolute value of the
volume of its environment with a coalescence model.
The 'charm' coalescence  volume 
would reflect partly  the QGP hot spot volume 
and partly the hadronic source volume 
from which hadrons with charm and anticharm can also
 form a $J/\Psi$.
If
 the  absolute yields per collision of
$J/\Psi$ and $D\overline{D}$ as a function of N, needed for this
calculation  would be published  by NA50, the
 charm coalescence volume could be calculated.  
\\ 

\noindent 
Figure \ref{jpsitoddbar_to_nto045} 
suggests that the coalescence picture could hold
for the full N range of Pb+Pb collisions up to N=380.
Obviously, it would be better to use the $D \overline{D}$ data themselves 
instead of the N parametrization, if high enough statistics would be
 available.
\\

\noindent
On the other hand, if the multiplicity of charm quarks
is high enough that often more than 1 charm quark pair
per event with charm is produced, then
it is the ratio $(J/\Psi)/(D\overline{D})^2$
which is expected to be inversely proportional to the volume 
of the charm source (this is exactly the case if $d$ coalescence out of
$p$ and $n$ is investigated in a baryon rich source).
The N dependence of the $(J/\Psi)/(D\overline{D})^2$
ratio, which would be relevant in the above discussed case
is
N$^{ (\alpha=-2.26 \pm 0.48) }$, $(\chi^2/DOF$=2.5, DOF=7)
respectively
N$^{ (\alpha=-3.1 \pm 0.24) }$, $(\chi^2/DOF$=1.2, DOF=6)
if the first point is not fitted.
The question on the absolute multiplicity of charm in
nuclear reactions, 
 should be answered by experiment.
\\

\subsection{The L dependence of the $(J/\Psi)/D\overline{D}$ ratio in nuclear collisions}
\vspace{0.5cm}

\vspace{0.4cm}
\noindent
{\bf Calculation details}
\vspace{0.2cm}

\noindent
The two distributions of $(J/\Psi)/D\overline{D}$ ratio
 for S+U and Pb+Pb collisions
in figure \ref{jpsitoddbar_vs_n_pbpb_su},
 are measured at different energies,
 and therefore they 
cannot be compared in terms of their absolute yields
but only with respect to their shapes.
\noindent
In order to compare their absolute yields, the data from figure 5 of 
\cite{phys_lett_b410_97_337} will be used.
There the $(J/\Psi)/DY$ ratio
in p+A, S+U and Pb+Pb collisions
 is shown as a function of L, all  normalised
to the same energy ($\sqrt{s}$=19 GeV) 
and corrected for the isospin dependence
of DY production.
The parameter L, is the length that the
 $J/\Psi$ traverses through nuclear matter.
In order to convert  figure 5 of \cite{phys_lett_b410_97_337}
to the $(J/\Psi)/D\overline{D}$ ratio as a function
of the L parameter, the $(J/\Psi)/DY$ ratio data points have been 
divided by the E factor 
as described in equation (3).
The correlation of the L parameter with N and b for Pb+Pb 
collisions, has been estimated using the theoretical calculation
of \cite{jyo_private}.
\\

\vspace{0.4cm}
\noindent
{\bf Results and discussion}
\vspace{0.2cm}

\noindent
The L dependence of the $(J/\Psi)/D\overline{D}$ ratio in 
arbitrary units in p+A, S+U and Pb+Pb collisions calculated here,
 is shown in figure \ref{jpsitody_vs_l}, 
together with the $(J/\Psi)/DY$ ratio
 published in \cite{fleuret,phys_lett_b410_97_337}.
The closed points show the $(J/\Psi)/D\overline{D}$ ratio in S+U and
Pb+Pb collisions extracted as indicated in equation 3.
The open squares and circles show the $(J/\Psi)/DY$ ratio in
S+U and Pb+Pb collisions from \cite{fleuret,phys_lett_b410_97_337}.
The open stars, show both the L dependence of the
$(J/\Psi)/DY$ as well as the L dependence
 of the $(J/\Psi)/D\overline{D}$ in p+A collisions
which are the same,
since the factor E has the value 1 for the latter.
\\

\noindent
The $J/\Psi$ over the $D\overline{D}$ production
investigated as a function of the volume through which the
 $J/\Psi$ traverses (V $\sim$ L$^3$)
is suppressed as compared to the shape of the exponential fit
 going through the $(J/\Psi)/D\overline{D}$ p+A data,
in both the S+U and Pb+Pb collisions at all $L$ points
(figure \ref{jpsitody_vs_l}).
\\

\noindent
The energy density of the lowest S+U point has been  
estimated to be $\sim$ 1.1 GeV/fm$^3$ \cite{cernep_2000_13},
which is comparable to the predicted critical energy density
for the QGP phase transition of $\sim$ 1 GeV/fm$^3$. 
A similar energy density of 1.2 GeV/fm$^3$ 
has been estimated \cite{cernep_2000_13} to be reached in the most 
peripheral Pb+Pb collisions measured by NA50.
\\

\noindent
In the following we investigate the initial energy density, 
rather than only the volume of the particle source (V $\sim$ $L^3$),
 as a critical parameter
for the appearance of the QGP phase transition.
\\

\section{The $\epsilon$ dependence of charm 
and strangeness in nuclear collisions}
\vspace{0.5cm}

\vspace{0.4cm}
\noindent
\subsection{\bf Charm}
\vspace{0.2cm}
\vspace{0.2cm}

\vspace{0.4cm}
\noindent
{\bf Calculation details}
\vspace{0.2cm}

\noindent
We estimate here
the $(J/\Psi)/ D \overline{D}$ ratio as a function of the energy density
$\epsilon$. 
For this purpose we use part of the data shown in
figure 7 in \cite{cernep_2000_13}.
There  the ratio of $((J/\Psi) /DY)_{measured}$ over the
 $( (J/\Psi) /DY)_{expected}$ is shown.
The '$( (J/\Psi) /DY)_{expected}$', is taken to be the exponential 
fit seen in figure \ref{jpsitody_vs_l}, which represents
 the 'normal' $J/\Psi$ dissociation 
 (i.e. understood without invoking QGP formation).
Dividing these data points by $N^{0.45 \pm 0.11}$, 
and  normalising the distribution of S+U and Pb+Pb points
 to the p+p and p+A data as in figure \ref{jpsitody_vs_l},
 we estimate the $( (J/\Psi) / D \overline{D})$ ratio over
  the
 expectation expressed by the above mentioned exponential curve,
which fits the 
$( (J/\Psi) / D \overline{D})$ data points for
p+p, p+d and p+A collisions.
\\

\vspace{0.4cm}
\noindent
{\bf Results and discussion}
\vspace{0.2cm}

\noindent
The result of this calculation  is shown in figure \ref{jpsi_vs_ed}
in logarithmic scale and in figure \ref{jpsi_k} (a) in linear scale.
It demonstrates a deviation of the $ (J/\Psi) / D \overline{D}$
ratio both in S+U and Pb+Pb collisions,
 from the p+p and p+A expectation curve,
occuring above $\epsilon$ $\sim$ 1 GeV/fm$^3$.
The 
logarithmic scale is shown to reveal  small changes in the slope of
the $( (J/\Psi)/ D \overline{D})$ distribution as a function of $\epsilon$,
appearing at $\epsilon$ $\sim$ 2.2 and 3.2 GeV/fm$^3$.
\\

\vspace{0.4cm}
\noindent
\subsection{\bf Strangeness}
\vspace{0.2cm}
\vspace{0.2cm}

\noindent
Figure \ref{jpsi_k} compares the two QGP signatures
of $J/\Psi$ suppression and of strangeness enhancement.
For this purpose we
represent all data points as a function of the estimated
energy density.
Note that the energy density as critical scale variable,
 has the advantage that unlike the temperature, 
it is defined irrespective of whether equilibrium is reached in the
collisions studied.
\\

\noindent
Figure  \ref{jpsi_k} (b) shows the multiplicity of kaons per event 
($K^+$, but also some $K^0_s$ data scaled to $K^+$ are shown)
divided by the effective volume of the particle source at thermal freeze out
in the center of mass frame, as a function of the initial energy density.
The effective volume represents the part of the real source volume, 
within which pions are correlated with each other (the so
 called 'homogenity' volume in the literature \cite{wiedemann_heinz_review}).
The effective volume is smaller than but proportional to the real source volume.
For a more precise calculation of the freeze out source volume
a detailled model is needed.
Here we estimate the effective volume at thermal freeze out $V_{thermal}$ 
based on measurements.
 The (smaller) effective volume at chemical freeze out $V_{chemical}$,
 is not experimentally measured, we give however
 an estimate of the ratio $V_{chemical}/V_{thermal}$.
Note that we compare the kaon data without rescaling for the different energy
between AGS and SPS\footnote{
 The  total $K^+$ multiplicity in p+p collisions,
 increases by a factor of $\sim$ 5, from 11.1 to 158 GeV per nucleon
 \cite{rossi}.}.
\\

\vspace{0.4cm}
\noindent
{\bf Calculation details}
\vspace{0.2cm}

\noindent
The effective volume V of the particle source has been estimated
in the center of mass frame,
 assuming cylindrical shape of the source:
\\

$ \ V \ = \  (\pi \cdot \ Radius_{cylinder}^2) \ \cdot \ Length_{cylinder} $
$\rightarrow$ 
\\

$ \ V \ = \  (\pi \ \cdot \ 4 \cdot \ R_{side}^2) \ \cdot \ (\sqrt{12} \ \cdot \ \cdot R_{long}) $
\\

\noindent
where $R_{side}$  is a measure of the transverse radius and 
$R_{long}$ is a measure of the longitudinal radius 
of the particle source, and the factors $4$ and $\sqrt{12}$ arise
from the definition of $R_{side}$, $R_{long}$ \cite{wiedemann_heinz_review}.
The $R_{side}$ and $R_{long}$
 values for central Au+Au collisions at 10.8 A GeV and
for central
 Pb+Pb collisions at 158 A GeV have been taken from \cite{ganz_qm99}. 
We dont use the more elaborated estimation
of the homogenity volume given in \cite{urs_wiedemann_qm99},
because the $R_{ol}$ component is not given in \cite{ganz_qm99}.
Based on the data of \cite{ganz_qm99}
we estimated the effective volume of the source at thermal freeze out
in central Au+Au collisions at 10.8 A GeV ($V \sim 1949 fm^3$)
and central Pb+Pb collisions at 158 A GeV ($V \sim 6532 fm^3$).
The effective volume increases by a factor of 3.35 from AGS to SPS energy.
\\

\noindent
Based on the temperature at thermal and chemical freeze out
which has been estimated from measurements using thermal
models \cite{t}, and the above estimated volumes at thermal freeze out,
we can further estimate the volume at chemical freeze out.
For this we assume that the relation $V \sim  T^{-3}$, which holds
in the universe
for massless particles in thermal equilibrium and for adiabatic expansion 
 \cite{kolb_turner},
holds approximately for heavy ion collisions at AGS and SPS energy.
Then from the temperature values at thermal and chemical freeze out
given in \cite{t} averaged over all models,
 we find that the ratio
$V_{chemical}/V_{thermal}$(AGS Si+Au 14.6 A GeV) =0.45
and
$V_{chemical}/V_{thermal}$(SPS Pb+Pb 158 A GeV) = 0.28.
Using the volume at chemical freeze out as estimated above,
 would stretch the $K/V$ ratio in figure \ref{jpsi_k} (b)
 between SPS and AGS, by a factor $\sim$ 1.6 apart.
We dont  use these values in figure \ref{jpsi_k}, because 
the above calculation is model dependent, e.g. the assumption of 
massless particles is not met, while the assumption of thermal equilibrium
may not be true.
\\

\noindent
The ratio K/N is expected to be proportional to
 the number density of kaons $\sim$ K/V, (V=volume),
 assuming V $\sim$ N (for justification of this assumption 
see footnote 3, \cite{ganz_qm99} and \cite{baker_qm99}).
Based on this expectation, we estimated here the K/V ratios
from the K/N ratios,  by normalising the K/N
ratios to the K/V value of the most central Au+Au events of E866
respectively Pb+Pb events of NA49, for which the value of the volume
has been estimated above.
\\

\noindent
The kaon data from Au+Au collisions at 11.1 A GeV (E866 and E802 experiments) 
 \cite{k_ags}
and from Pb+Pb collisions at 158 A GeV (NA49 experiment) \cite{sikler}
are kaon multiplicities extrapolated to full acceptance.
Therefore NA49 and E866 data are absolutely normalised.
We estimated  K/N from the NA49 experiment using
 the kaon multiplicities from
\cite{sikler} and the
 number of wounded nucleons from \cite{seyboth_hi} as 
available\footnote{
We take N equal to
 the number of wounded nucleons for NA49, because it is 
used in all other experiments presented here, and allows for a
straightforward geometrical interpretation of N.},
otherwise we used the N estimated from the experimental
baryon distribution  \cite{sikler}. 
\\

\noindent
The data from NA52 \cite{na52_centr} and WA97 \cite{k_wa97}
have been measured in a small phase space acceptance
and have been scaled here arbitrarily, in order to match
the NA49 data in figure \ref{jpsi_k}.
This scaling 
is justified since all NA52, NA49 and WA97 measurements
are kaons produced in Pb+Pb collisions at 158 A GeV,
and 
'extrapolates' the NA52 and WA97 data to
the NA49 full acceptance multiplicities,  allowing for comparison of
the shapes of the distributions.
It is assumed, that the N and $\epsilon$ dependence of kaons 
does not change significantly with the phase space acceptance.
\\

\noindent
In order to calculate the energy density we have performed the following steps.
The energy density for all colliding systems has been estimated
 using the Bjorken formula \cite{bjorken}
and data given in \cite{k_ags,k_ed,na49_et}.
The transverse radius of the overlapping region of the colliding nuclei
is found as: $R_{trans} = 1.13 \cdot (N/2)^{1/3} $, where N is
the total number of participant nucleons.
The formation time was taken 1 fm/c \cite{bjorken}.
\\

\noindent
For the E866 experiment, lacking $E_T$ values, (but with measured
 E$_{forward}$), we used instead of $(d E_T/d \eta)_{ycm}$, ($ycm$=midrapidity)
the total energy of the nucleons participating in the collision
($E_{tot,part}=N_{projectile,participants} \cdot E_{beam}$),
assuming the proportionality $(d E_T/d \eta)_{ycm} \sim E_{tot,part}$,
and we further normalised the  results in such a way that
 the maximum energy density of our estimate  matches
the absolute value of the maximal achieved energy density in the most
central Au+Au events at this energy of 1.3 GeV/fm$^3$,
given in \cite{k_ed}.
\\

\noindent
NA52 measures $E_T$  near midrapidity (y $\sim$ 3.3).
 These values were used to estimate the energy density
  and the results have been
normalised to the maximum energy density reached in Pb+Pb collisions
at the same centrality of $\epsilon_{max}$=3.2 GeV/fm$^3$,
 extracted by NA49 in \cite{na49_et}.
Parametrizing the dependence of the energy density on the
number of participants found from the NA52 data as described above,
 we  estimated the energy density corresponding to the N values of the WA97
and the NA49 kaon measurements, given in \cite{k_wa97,sikler}.
Data from S+S collisions taken from \cite{sikler} and \cite{ed_ss,na49_et}
are also shown.
\\

\noindent
To estimate the systematic error on the energy
density found with the above methods, we calculated the  energy density
in Pb+Pb collisions at 158 A GeV, using the VENUS 4.12
\cite{venus412} event generator.
We estimated with VENUS the $(d E_T/d \eta)_{ycm}$ at $ycm$=midrapidity and 
the number of participant nucleons and used them to find the
energy density from the Bjorken formula \cite{bjorken}.
The deviation of the energy density calculated with VENUS
 $(d E_T/d \eta)_{ycm}$
 from the energy density found using the NA52 transverse energy
 measurements is $\leq$ 30\% of the latter.
The deviation of the energy density calculated with VENUS $d E_T/d \eta$
  from the energy density found using the total energy of the participant
nucleons and of the newly produced particles, 
(which is similar to
 the method used to estimate the energy density for the AGS data),
over the latter energy density,
  is at the same level. 
\\

\noindent
In this context, it appears important for a more precise comparison of data
as a function of $\epsilon$,
 that experiments publish together with the number of participants also
the $d E_T/d \eta$ at midrapidity for each centrality region,
 for both nucleus+nucleus and for p+p collisions,
 estimated by models or measured if available (e.g. in NA49). 
\\

\vspace{0.4cm}
\noindent
{\bf Results and discussion}
\vspace{0.2cm}

\noindent
Figure \ref{jpsi_k} (b) suggests that
kaons below $\epsilon$ $\sim$ 1 GeV/fm$^3$ did not reach equilibrium, while
this seems to be the case above.
Indeed kaons produced in Au+Au collisions at 11.1 A GeV
 \cite{k_ags} and in very peripheral Pb+Pb collisions at 158 A GeV 
\cite{mypadova98,myqm99,na52_centr},
increase faster than linear with N, indicating non thermal
kaon production, while they increase nearly proportional to N
above $\epsilon$ $\sim$ 1 GeV/fm$^3$ \cite{myqm99,na52_centr,k_wa97}.
The connection of strangeness equilibrium and
 the QGP phase transition has been
 discussed e.g. in \cite{rafelski_mueller_koch}.
There it is shown, that
strangeness in heavy ion collisions 
is expected to reach equilibrium values
if the system runs through a QGP phase, while this is 
less probable in a purely hadronic system. 
\\

\noindent
Figure \ref{jpsi_k}
demonstrates that both the $J/\Psi$ and kaon production
exhibit a dramatic change above the energy density
of $\sim$ 1 GeV/fm$^3$.
While the equilibration of strange particles as suggested by their
$\sim N^1$ dependence above 1 GeV/fm$^3$,
could in principle
 also be due to equilibrium reached in a hadronic environment,
the combined appearance of this effect and of the
$ (J/\Psi) / D \overline{D}$ suppression at the same
energy density value is a striking result, indicating
a change of phase above $\epsilon_c$=1 GeV/fm$^3$.
\\

\noindent 
The expectation for the shape of the $J/\Psi$ suppression
as a function of energy density are three succesive drops of the
$J/\Psi$ \cite{satz_qm99,cernep_2000_13}; 
a drop by $\sim$ 8$\%$ 
\cite{gerschel_paper}
due to $\psi^{'}$ dissociation,
a drop by $\sim$ 32$\%$ 
\cite{gerschel_paper}
 due to the $\chi_c$ dissociation 
and 
a drop by $\sim$ 100$\%$ due to the $J/\Psi$ dissociation.
All these without taking into account regeneration of
$J/\Psi$ through other processes.
These can be e.g. coalescence of
charm quarks or $J/\Psi$ not travelling through the plasma. 
The $\psi^{'}$ feeds only 8$\%$ of the total $J/\Psi$'s and can
therefore hardly  be observed as a break in the $J/\Psi$ production.
\\

\noindent
The absolute value of the energy density
 $\epsilon$ and therefore of the N values  at which
 these changes could be observed is not exactly given by the models.
The
 critical energy densities for the dissociation
of the states $\Psi^{'}$, $\chi_c$ and $J/\Psi$
could even be so near to each other that no clear multistep behaviour
is seen in $ (J/\Psi) /D \overline{D}$.
\\

\noindent
Figure \ref{jpsi_k} suggests that the breaks in the $ (J/\Psi)/D \overline{D}$ ratio
at $\epsilon$ $\sim$ 2.2 and 3.2 GeV/fm$^3$, are  less dramatic than the change above
 $\epsilon$ $\sim$ 1 GeV/fm$^3$. 
Therefore, all bound $c \overline{c}$ states 
  could be dissociated at similar energy densities, which lye near 1 GeV/fm$^3$.
\\

\noindent
Alternatively, the $\psi^{'}$ and the
 $\chi_c$ could dissociate
 above $\epsilon$ $\sim$ 1 GeV/fm$^3$
and the dissociation of the $J/\Psi$ 
could start at $\epsilon$=2.2 GeV/fm$^3$, if we interpret
the change
in the $ (J/\Psi) / D \overline{D}$ ratio, below and 
above $\epsilon$=2.2 GeV/fm$^3$, as a step behaviour.
In
 this context, the steep drop of the $ (J/\Psi)/D \overline{D}$ ratio
in the  bin(s) of largest N  
(figures \ref{jpsitoddbar_to_nto045}, \ref{jpsi_vs_ed}, \ref{jpsi_k}),
 cannot be interpreted in a natural way.
 The steps  of $ (J/\Psi) /D \overline{D}$ remain to be established
through a  direct measurement of $J/\Psi$ and $D \overline{D}$ absolute
yields as a function of ($E_T$, N, $\epsilon$).
\\

\noindent
In the above discussed picture, three QGP  signatures
 appear in nuclear collisions at energy density 
larger than $\sim$ 1 GeV/fm$^3$:
\\
 a) $J/\Psi$ suppression
(figure \ref{jpsi_k} (a))
 --which could be due  to 
bound $c \overline{c}$ states dissociation--
\\
b) strangeness enhancement 
(figure \ref{jpsi_k} (b)),
possibly
due to equilibration of $s \overline{s}$ in QGP as opposed to hadrons,
\\
 c) 
the invariant mass m($e^+e^-$)  
excess at m below the $\rho$ mass \cite{ceres},
possibly due to a $\rho$ change \cite{rho} and/or
to increased production of the lowest mass glueball state in QGP
\cite{gb}.
\\
\noindent
This
 coincidence of QGP signatures, suggests a 
 change of phase at $\epsilon$ $\sim$ 1 GeV/fm$^3$
as expected \cite{lattice}.
\\

\noindent
From the above discussion  it follows, that a direct measurement
of  open charm production in nuclear collisions 
appears essential for the physics of the 
 Quark Gluon Plasma phase transition.
Furthermore, if enhanced over expectations, open charm in nuclear
collisions defies theoretical understanding.
\\

\section{Possibilities for future measurements}
\vspace{0.5cm}

\noindent
A measurement of open and closed charm production in Pb+Pb
collisions as a function of energy
 below the SPS top energy of $\sqrt{s}=17 $ GeV
 searching for the disappearance of the
 seen $J/\Psi$ suppression in central Pb+Pb
collisions at a certain $\sqrt{s}$, 
 could prove clearly the QGP phase transition.
Using the same nuclei at different $\sqrt{s}$
and looking only at central collisions, differences
due to different nuclear profiles
drop out.
No currently existing or planned experiment at SPS is 
however able to perform this measurement without major upgrades,
though one proposal (NA6i) could 
significantly improve the identification of open charm production through a
better determination of the decay vertex \cite{na6i}. 
An upgrade of NA50/NA6i, or
alternatively  a completely new experiment,
could possibly achieve this goal.
The study  could also  be performed at the Relativistic Heavy
Ion Collider (RHIC) using lower energy and/or
large and small nuclei, and in fixed target
experiments at RHIC favoured because of higher luminosity
as compared to the collider mode,  important 
for a low energy scan.
\\

\noindent
It would be also important (and easier than the above)
 to measure
the $J/\Psi$, $D \overline{D}$ and $DY$  absolute yields 
per collision, below $\epsilon$=1 GeV/fm$^3$,
by using the most peripheral (not yet investigated)
 Pb+Pb collisions or collisions of lighter nuclei 
 at the highest beam energy at SPS
($\sqrt{s}$ = 17,19 GeV).
\\

\noindent
An other  piece of information important for
the understanding of charm production in nuclear collisions
would be 
 the direct comparison of the $ (J/\Psi)/DY$ and the $(J/\Psi)/D \overline{D}$
ratios in nuclear collisions at $\sqrt{s} <$ 19 GeV
 and in $p +\overline{p}$ collisions at the Tevatron.
Tevatron reaches an energy density similar to or larger than
 the one estimated in very
 central S+S collisions at 200 A GeV \cite{na35}.
Therefore it would supply a comparison for these points
and a continuation of the absorption line
fitted through the p+p and p+A data measured by NA50
(figure \ref{jpsitody_vs_l}), or otherwise.
Differences  due to the change of
dominant production mechanisms of charm in $p \overline{p}$ collisions
as compared to A+B, p+p, can be accounted for theoretically.
A high $E_T$ cut could additionally help in
sorting out 'central' $p+ \overline{p}$ collisions.
This comparison should be done  possibly 
 in the very same dimuon mass region for all processes
(also DY)
e.g.
using Monte Carlo's  tuned to $p +\overline{p}$ Tevatron data.
\\

\noindent
This comparison would answer the question,
 if the energy density is indeed the only critical variable
for the appearance of
a thermalised QGP state with 3 effective flavours u,d,s,
 or whether there is also a critical volume (e.g. as measured by
the L variable:  V $\sim$ L$^3$).
Furthermore, at present the comparison of nuclear collisions 
to
p+p and p+A data is done at the same 
energy and not at the same energy density.
This issue is important, since if for example
 the energy density is the only critical scale
variable,
  the QGP should be formed also in elementary collisions like 
$p \overline{p}$ at a higher
beam energy and the same energy density.
\\

\noindent
Further it is important to search for thresholds in the production
of many particles e.g. $\Omega$, which was found to be enhanced
by a factor 15 above p+A data in  Pb+Pb collisions
at 158 A GeV \cite{omega} in the energy density region corresponding to 
the green stars in figure \ref{jpsi_k} (b). 
Similarly interesting would be
a measurement of the invariant mass of $e^+ e^-$ in low energy densities.
\\

\section{Conclusions}

\noindent
In this letter, consequences resulting from the viable possibility that the
dimuon invariant mass 
(m($\mu^+ \mu^-$)) enhancement, measured by the NA50 experiment
 in the intermediate mass region (IMR):
 $m(\phi) < m(\mu^+ \mu^-) < m(J/\Psi)$,
 in S+U and Pb+Pb collisions at $\sqrt{s}$ 19, 17 GeV,
reflects a $D\overline{D}$ enhancement over
expectations, are worked out.
\\

\begin{figure}[htb]
\begin{minipage}[t]{100mm}
\epsfxsize=250pt
 \hspace*{3.0cm}
\leftline{ \epsfbox{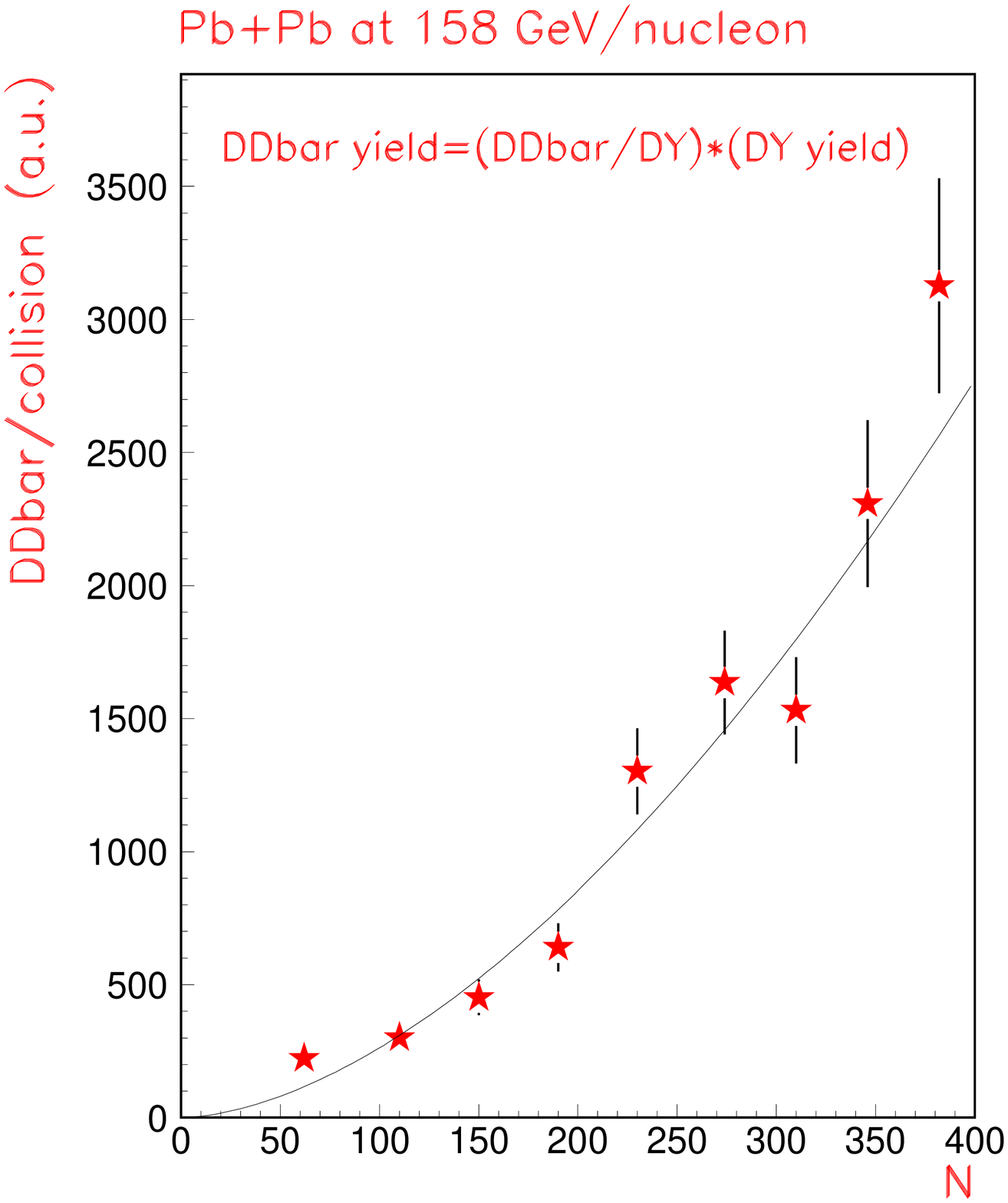} $\;\;\;\;\;\;\;\;\;$  }
\end{minipage}
\caption{$D \overline{D}$ yield per collision in arbitrary
units in Pb+Pb collisions at 158 A GeV,
as a function of the number of participating nucleons N.
The line shows the result of fitting the function $f=c N^{\alpha}$.
See text 
for the fit results.
}
\label{ddbaryield}
\end{figure}

\begin{figure}[htb]
 \hspace*{3.0cm}
\begin{minipage}[t]{100mm}
\epsfxsize=250pt
\leftline{ \epsfbox{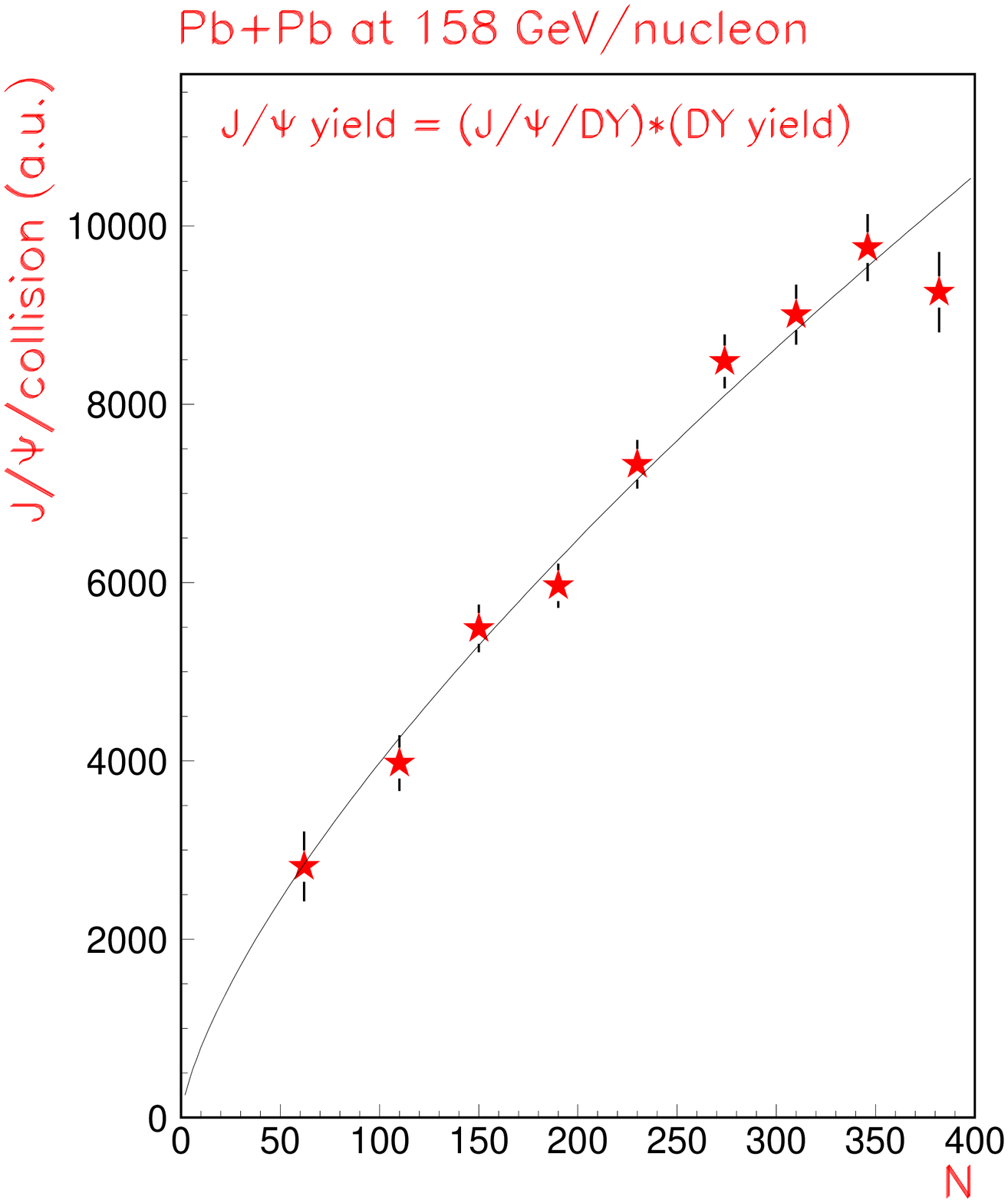} $\;\;\;\;\;\;\;\;\;$  }
\end{minipage}
\caption{$J/\Psi$ yield per collision in arbitrary
units in Pb+Pb collisions at 158 A GeV,
as a function of the number of participating nucleons N.
The line shows the result of fitting the function $f=c N^{\alpha}$.
See text 
for the fit results.
}
\label{jpsiyield}
\end{figure}

\noindent
 The dependence of the $J/\Psi$ and the $D\overline{D}$
yields per collision in Pb+Pb collisions 
on the mean number of participants has been estimated.
This dependence reveals the non thermal features of
 charm production at this energy.
The $\sim$ $N^{0.7}$ dependence of the $J/\Psi$ yield  
(figure \ref{jpsiyield})
suggests strong dissociation of $J/\Psi$ with higher centrality. 
The dissociation is stronger than the absorption seen in any
other hadron, e.g. $\overline{p}$ in Pb+Pb collisions.
The N dependence of the $D\overline{D}$ yield of   $N^{1.7}$
(figure \ref{ddbaryield})
 indicates
also 
non-thermal open charm  production at this energy, showing up
in an excess rather than reduction as compared to the
thermal expectation.
\\

\begin{figure}[htb]
 \hspace*{3.0cm}
\begin{minipage}[t]{100mm}
\epsfxsize=250pt
\leftline{ \epsfbox{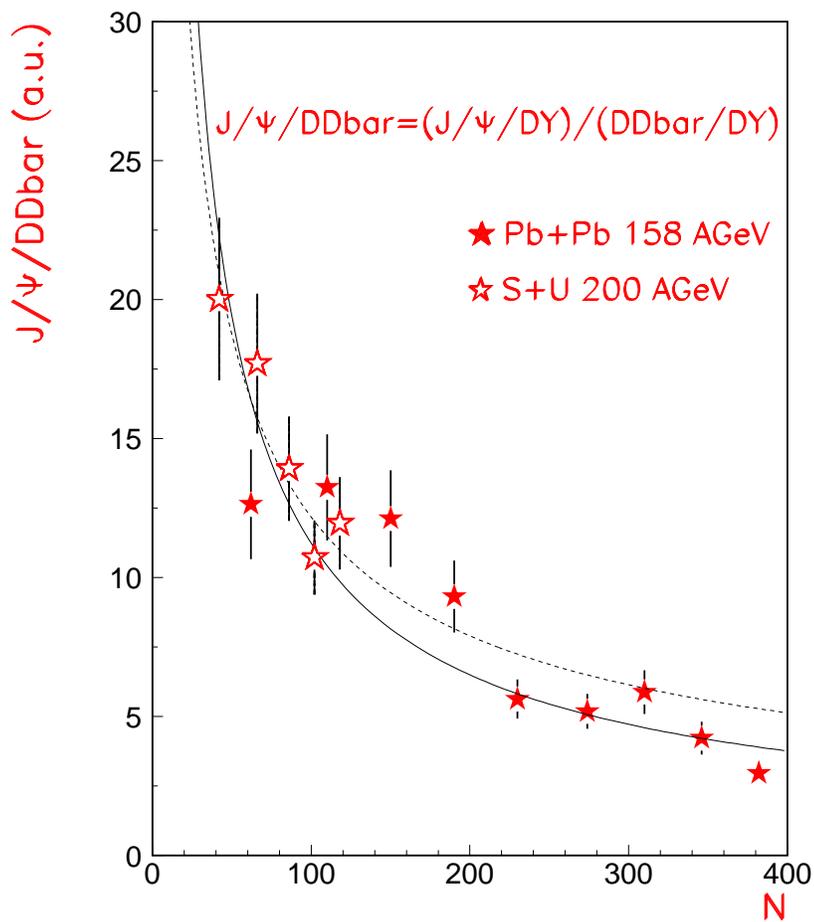} $\;\;\;\;\;\;\;\;\;$  }
\end{minipage}
\caption{Ratio of $J/\Psi$ to $D \overline{D}$ in
 arbitrary units in Pb+Pb collisions at 158 A GeV (closed points)
and in S+U collisions at 200 A GeV (open points),
as a function of the number of participating nucleons N.
The lines show the result of fitting the function $f=c N^{\alpha}$
to the S+U (line above) and to the Pb+Pb data points (line below).
See text 
for the fit results.
}
\label{jpsitoddbar_vs_n_pbpb_su}
\end{figure}

\begin{figure}[htb]
 \hspace*{3.0cm}
\begin{minipage}[t]{100mm}
\epsfxsize=250pt
\leftline{ \epsfbox{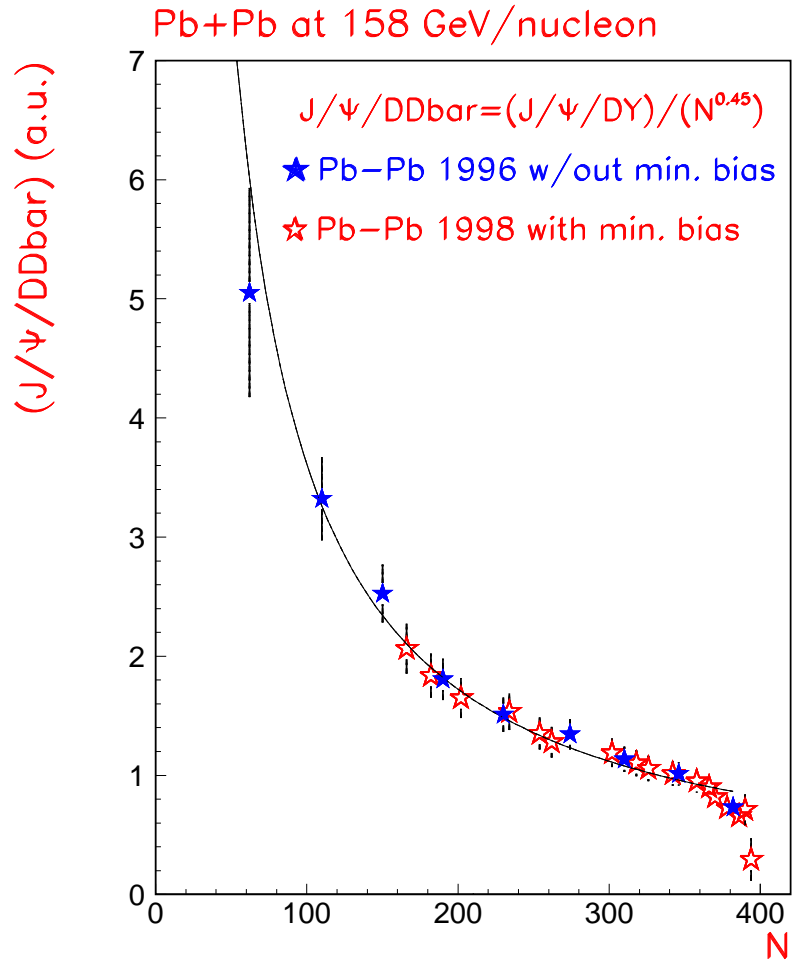} $\;\;\;\;\;\;\;\;\;$  }
\end{minipage}
\caption{Ratio of $J/\Psi/DY$ to $N^{0.45 \pm 0.11}$ 
 in Pb+Pb collisions at 158 A GeV
as a function of the number of participating nucleons N.
The 9 closed star points correspond to the N values at which
the $D \overline{D}$ yield was measured.
The lines show the result of fitting the function $f=c N^{\alpha}$
to the closed stars.
See text for the fit results.
}
\label{jpsitoddbar_to_nto045}
\end{figure}

\noindent
If
 the  dimuon excess observed by NA50 is  partly or solely due to open charm,
 it is appropriate to search for an anomalous suppression of
$J/\Psi$ as compared to the total open charm production, rather than
to the DY process.
We therefore investigated here
the $ (J/\Psi)/D\overline{D}$ ratio in Pb+Pb collisions
 and we find it to
decrease approximately as $\sim$ $N^{-1}$ 
 (figures \ref{jpsitoddbar_vs_n_pbpb_su}, \ref{jpsitoddbar_to_nto045}).
This is the N dependence expected for the $J/\Psi$ if 
it were completely dissociated in quark gluon matter
 and were later
dominantly formed through $ c \overline{c}$ quark coalescence, assuming
N $\sim$ volume of the $ c \overline{c}$ environment and charm quark
multiplicity of one in events with charm.
\noindent
In that case, based on coalescence arguments,
the $ (J/\Psi)/ D \overline{D}$ ratio
could  be  used to estimate the volume of the charm environment, which may
reflect partly the  size of the  quark gluon plasma.
This is probable under the assumption
 that the final measured $J/\Psi$ is dominated by the
$J/\Psi$ originating from $c \overline{c}$ pairs which travel through the
plasma volume, an assumption which may hold only for
large plasma volumes, i.e. for the most central collisions.
\\

\begin{figure}[htb]
\begin{minipage}[t]{100mm}
 \hspace*{3.0cm}
\epsfxsize=250pt
\leftline{ \epsfbox{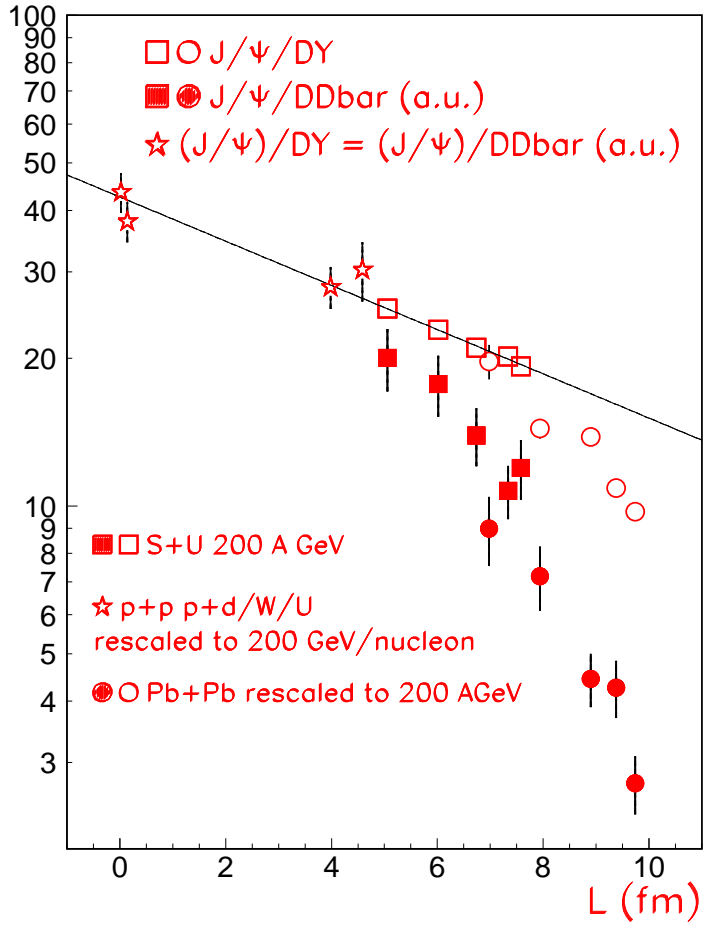} $\;\;\;\;\;\;\;\;\;$  }
\end{minipage}
\caption{
The open points show the ratio of $J/\Psi$ to Drell Yan ($DY$)  
in p+A, S+U and Pb+Pb collisions normalised to
200 A GeV, as a function of the path of $J/\Psi$ through nuclear matter
L \protect\cite{phys_lett_b410_97_337,fleuret}.
The closed points show the ratio of $J/\Psi$ to $D \overline{D}$ 
  in S+U and Pb+Pb collisions at 200 A GeV estimated here.
The open stars show the L dependence of the  $J/\Psi/DY$ and the
$J/\Psi/D \overline{D}$ ratio in p+A collisions.
All $J/\Psi / D \overline{D}$ ratios shown in this figure are in common
arbitrary units.
}
\label{jpsitody_vs_l}
\end{figure}

\begin{figure}[htb]
\begin{minipage}[t]{100mm}
 \hspace*{3.0cm}
\epsfxsize=250pt
\leftline{ \epsfbox{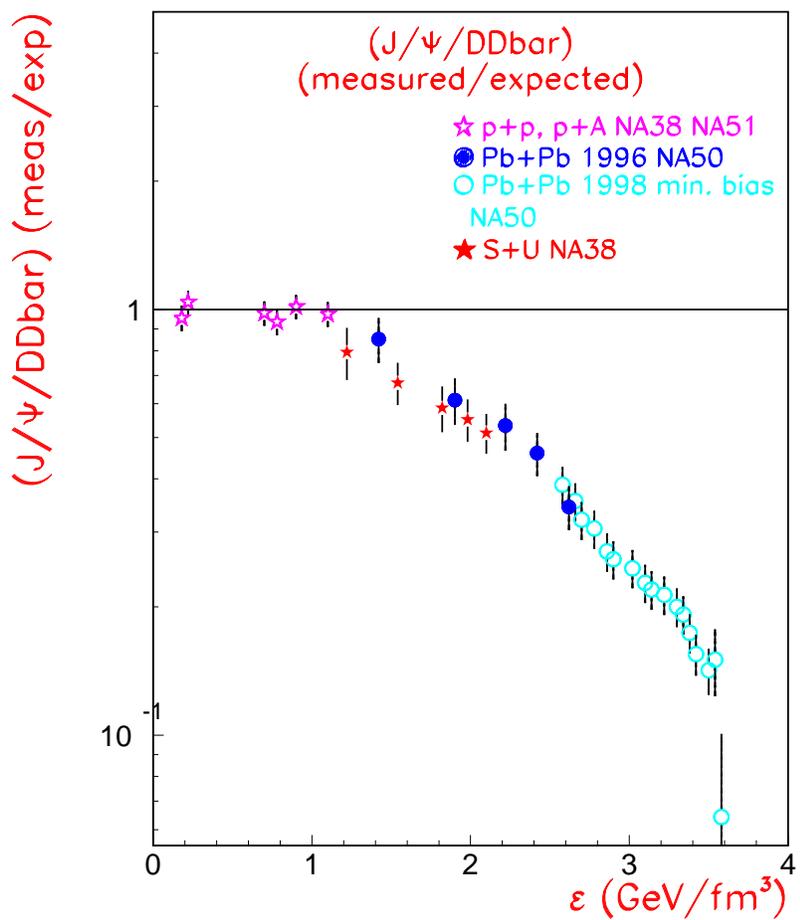} $\;\;\;\;\;\;\;\;\;$  }
\end{minipage}
\caption{
The $J/\Psi/D \overline{D}$ (measured/'expected') ratio is shown
as a function of the initial energy density ($\epsilon$) achieved in the
collisions investigated.
}
\label{jpsi_vs_ed}
\end{figure}

\noindent
A further consequence of a possible open charm enhancement is 
 that the $J/\Psi$ over the $D\overline{D}$  ratio
appears to be suppressed already in
S+U collisions as compared to p+A collisions,
 unlike the $ (J/\Psi)/DY$ ratio (figure \ref{jpsitody_vs_l}).
The $\psi^{'}/D \overline{D}$ ratio
 would also be additionally suppressed as compared
to the $\psi^{'}/DY$ in both S+U and Pb+Pb collisions. 
These phenomena could be interpreted as onset of dissociation
of bound charm states 
above energy density $\epsilon$ $\sim$ 1 GeV/fm$^3$.
\\

\begin{figure}[htb]
\begin{minipage}[t]{70mm}
\epsfxsize=200pt
\leftline{ \epsfbox{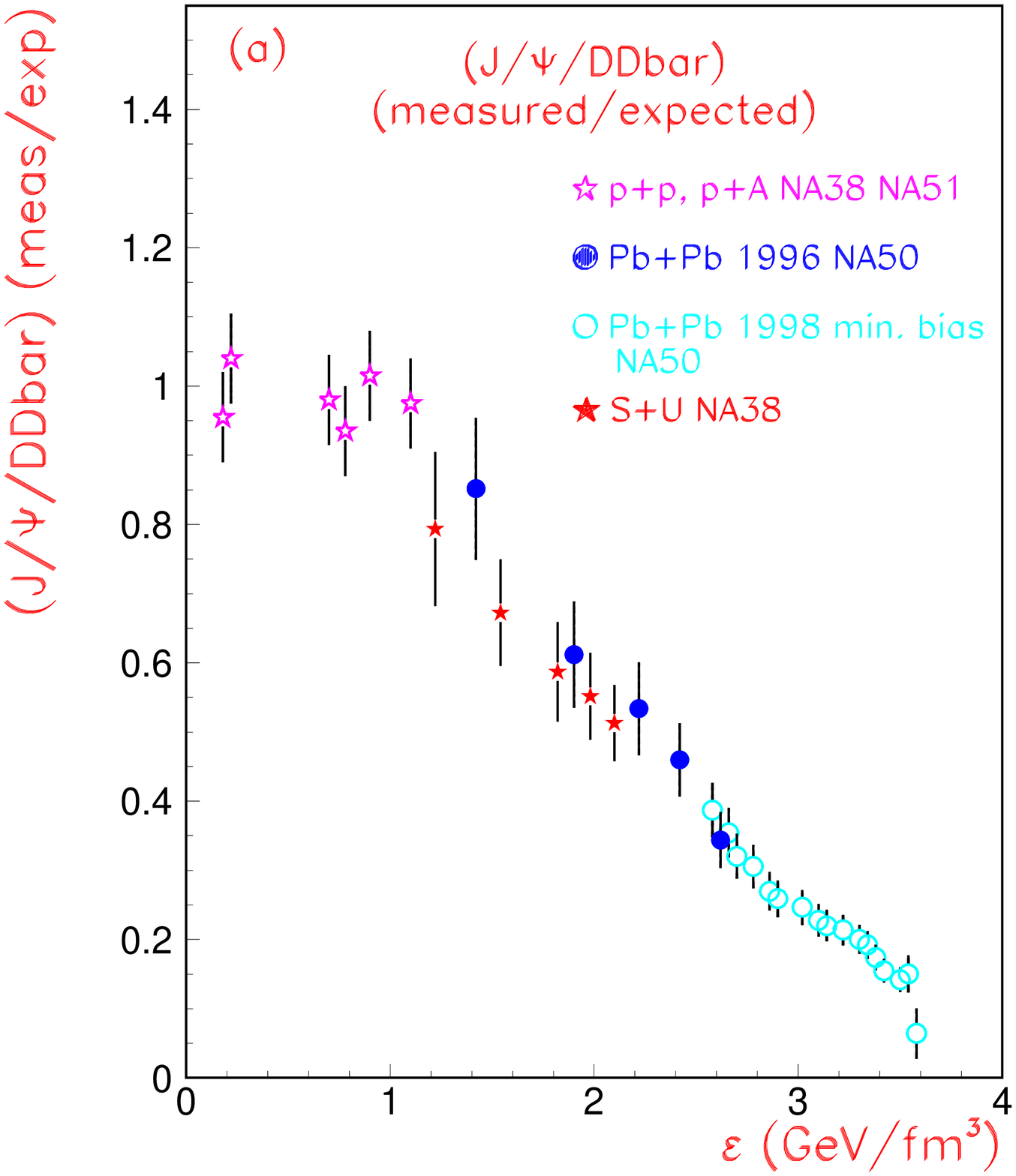} $\;\;\;\;\;\;\;\;\;$  }
\end{minipage}
\begin{minipage}[t]{70mm}
 \hspace*{1.2cm}
\epsfxsize=200pt
\leftline{ \epsfbox{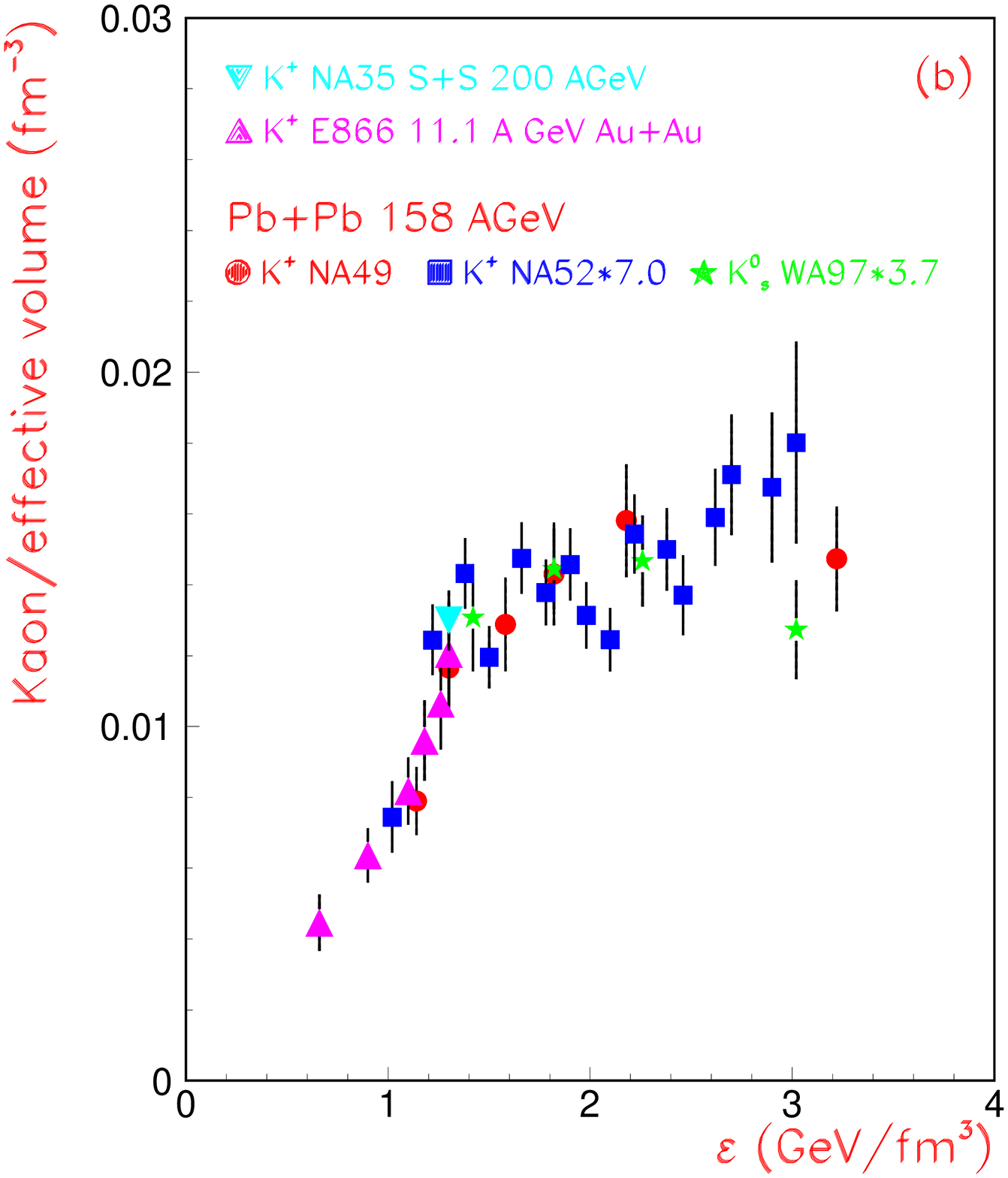} $\;\;\;\;\;\;\;\;\;$  }
\end{minipage}
\caption{
(a) The $J/\Psi/D \overline{D}$ (measured/'expected') ratio is shown
as a function of the initial energy density ($\epsilon$) achieved in the
collisions investigated.
(b)
The kaon ($\sim$ $K^+$) multiplicity 
 over the effective volume
($V= (\pi \cdot  4 \cdot  R_{side}^2) \cdot (\sqrt{12} \cdot R_{long}) $)
of the particle source at thermal freeze out, in the center of mass frame,
is shown as a function of 
the initial energy density ($\epsilon$).
The above effective volume
 is smaller than the real source volume but proportional to it.
}
\label{jpsi_k}
\end{figure}

\noindent
We estimated and compared the dependence of the $ (J/\Psi)/D\overline{D}$  
ratio and of the kaon multiplicity per volume
 (assuming K/N $\sim$ K/V = kaon number density)
in several collisions and $\sqrt{s}$
as a function of the initial energy density.
We find that both the kaon number density and the ratio
$ (J/\Psi)/ D \overline{D}$ 
exhibit dramatic changes at the energy density
of 1 GeV/fm$^3$, as demonstrated in figure \ref{jpsi_k}.
This is the main result of this paper.
\\

\noindent
It follows that three major QGP signatures
($s\overline{s}$ enhancement, $\rho$ changes and 
$J/\Psi$ suppression) all appear  
above the energy density of $\sim$ 1 GeV/$fm^3$, which
 is the critical energy density for the QGP phase transition
according to lattice QCD. 
\\

\noindent
This  discussion underlines the importance of a direct measurement of open 
charm production in nuclear collisions,
and of other experimental investigations proposed in section 5,
 for the understanding of ultrarelativistic nuclear
reactions and the dynamics 
 of the Quark Gluon Plasma phase transition.

\vspace{1.0cm}

\noindent
Acknowledgments\\

\noindent
I would like to thank Prof. P. Minkowski, Prof. K. Pretzl, Prof. U. Heinz and
Prof. J. Rafelski for stimulating discussions, 
and 
 Dr. C. Cicalo, Dr. O. Drapier, Dr. C. Gerschel, 
 Dr. E. Scomparin, Dr. P. Seyboth, Dr. F. Sikler, Dr. U. Wiedemann,
and especially Dr. J.Y. Ollitrault 
for clarifying discussions on their data and/or for 
communicating their results to me.

\newpage

\end{document}